# A Variational Approach to Extracting the Phonon Mean Free Path Distribution from the Spectral Boltzmann Transport Equation

Vazrik Chiloyan[a], Lingping Zeng[a], Samuel Huberman[a], Alexei A. Maznev[b], Keith A. Nelson[b], Gang Chen[a]*

[a]Department of Mechanical Engineering, Massachusetts Institute of Technology, Cambridge, Massachusetts 02139, USA
[b]Department of Chemistry, Massachusetts Institute of Technology, Cambridge, Massachusetts 02139, USA

**Abstract**

The phonon Boltzmann transport equation (BTE) is a powerful tool for studying non-diffusive thermal transport. Here, we develop a new universal variational approach to solving the BTE that enables extraction of phonon mean free path (MFP) distributions from experiments exploring non-diffusive transport. By utilizing the known Fourier solution as a trial function, we present a direct approach to calculating the effective thermal conductivity from the BTE. We demonstrate this technique on the transient thermal grating (TTG) experiment, which is a useful tool for studying non-diffusive thermal transport and probing the mean free path (MFP) distribution of materials. We obtain a closed form expression for a suppression function that is materials dependent, successfully addressing the non-universality of the suppression function used in the past, while providing a general approach to studying thermal properties in the non-diffusive regime.

*Corresponding author: gchen2@mit.edu

The Boltzmann transport equation (BTE) is widely used in analyzing heat transfer at length scales and time scales for which Fourier's law breaks down. In particular, there has been a growing interest recently in developing numerical and analytical solutions to the BTE to model thermal transport in phonon spectroscopy experiments [1–11] to extract phonon mean free path (MFP) distribution. The thermal conductivity accumulation function has been utilized as an elegant metric for understanding which MFP phonons contribute predominantly to thermal transport in a material [12–14]. Various experimental tools such as time-domain thermoreflectance [4,6,8,15,16] (TDTR), frequency-domain thermoreflectance [17,18] (FDTR), and transient thermal grating [3,5,7,19] (TTG) techniques have been utilized extensively recently in order to probe and observe non-diffusive transport by using ultrafast time scales or ultrashort length scales and gain key insight into the material's MFP spectrum.

When the length scales in a system become comparable to the MFPs in a material, the effective thermal conductivity is reduced compared to its bulk, diffusive limit value [20,21]. A suppression function $S_\omega$ is used to quantify this reduction or suppression of thermal conductivity, defined as $k_{eff} = \frac{1}{3}\int_0^{\omega_m} C_\omega v_\omega \Lambda_\omega S_\omega \, d\omega$. The suppression function provides the ability to extend the notion of thermal conductivity beyond the diffusive regime in which it is defined from Fourier's law [20,22]. By utilizing the suppression function for a given experimental geometry, one can obtain the material's phonon MFP distribution from the experimentally measured thermal conductivity [5,6,22]. To obtain the effective thermal conductivity, the thermal signal from the experiment is fitted to the results of the Fourier law. The suppression function is calculated through modeling of the given experimental geometry with the BTE. However, one key assumption in this method is the universality of the suppression function, i.e. the ability to express the suppression function as $S_\omega = S(\Lambda_\omega / L)$ so that it depends only on the ratio of MFP to a characteristic length for a given experimental configuration, but not otherwise on the material properties. This assumption allows one to obtain effective thermal conductivities by solving for the suppression function from the gray BTE, i.e. the BTE equation with a single MFP [23]. This assumption has been shown to be not strictly valid in the past [23,24] and will be further shown in this work, with an approach that addresses this shortcoming.

The BTE is notoriously difficult to solve, especially for complex geometries, which presents difficulty in calculating the effective thermal conductivity of materials in a given experimental geometry. So far almost exclusively, numerical solutions are implemented that directly attempt to solve the BTE, and are then fitted to the Fourier solution to extract the effective thermal conductivity and corresponding suppression function for the experimental geometry. Experimental methods have also been utilized that rely on first-principles material property data to obtain a calibrated suppression function [6]. The key insight in our work here is to utilize the temperature distribution obtained from the Fourier diffusion equation directly in the BTE for the given experimental geometry to facilitate, hence significantly simplify, its solution. Furthermore, we develop a variational approach to yield solutions to the spectral BTE. By utilizing the temperature field derived from the Fourier solution and the variational method, we obtain solutions that are both simple yet can reproduce the exact numerical results from the BTE in terms of obtaining the effective thermal conductivity. Our approach provides a more direct, universal methodology for extracting the effective thermal conductivity and corresponding suppression function to enable the extraction of intrinsic material properties such as the phonon MFP distribution from non-diffusive experiments.

The variational approach utilized here for the BTE is analogous to the variational method in quantum mechanics, used for improving one's trial solution for the ground state energy of a given system [25,26]. The variational principle has been applied to the BTE previously in calculating the cross plane heat flux in a thin film [27]. Allen [27] utilized a specific error metric, one that tries to best enforce uniform heat flux through the slab to ensure energy conservation, to approximately calculate the thermal flux between a hot wall and cold wall. Furthermore, variational techniques have been applied to solving the BTE at the initial stage of the partial differential equation itself [28,29]. In this approach, first described by Ziman [30], the partial differential equation for the phonon distribution function is solved utilizing the variational principle, and the variational parameter is calculated by optimizing the entropy. In this work, we develop the application of the variational principle upon the integral equation for the temperature profile, derived from the BTE, and solve for the variational parameter by minimizing the residual error in the equation. Although we anticipate that this approach can be applied also directly to the BTE, the technique developed here is applied at the stage of the temperature equation for two reasons. One being that analytically solving the BTE up to the temperature equation decreases inaccuracies that can build up from utilizing approximations earlier on in the solution. Second, the temperature equation allows for the direct utilization of the Fourier solution as the trial function, with effective thermal conductivity (or other properties such as interfacial resistance or boundary temperature slip) as the parameters used to minimize the error of the variational trial function.

In this work, we apply the variational technique for the one-dimensional TTG experimental geometry as an example. In the TTG experiment, two laser beams are crossed in order to generate a sinusoidal heating profile on a sample, with a spatial periodicity of length $\lambda$. Once heated, the sample is allowed to relax and the thermal decay profile is measured to yield information about the transport within the material. At grating periods on the order of micrometers, non-diffusive transport has been observed [3,7,19]. Given the success of this experiment in probing non-diffusive transport and the opportunity to yield MFP data using reconstruction techniques that have been developed [31], the ability to model this experiment is critical. Furthermore, the relative simplicity of the geometry makes it more accessible for theoretical modeling.

The TTG in the one-dimensional case has been studied in a two-fluid framework, and with simplifying assumptions about the scattering of high and low frequency phonons, an analytical suppression function has been calculated [19] and utilized in MFP reconstruction [31], but there is a concern that this model is only valid at the onset of non-diffusive transport. Collins *et al.* solved the problem with a numerical approach to obtain the exact solution both in the gray case as well as the full spectral case for the BTE for Si and PbSe [23]. Deviation of the two-fluid model from the exact numerical solution was shown for PbSe [23]. Hua and Minnich obtained the Fourier transform of the thermal decay analytically, and were able to recover the two-fluid model suppression function in the weakly non-diffusive limit [24]. However, there is no closed form expression for the thermal decay rate $\gamma$ and the suppression function $S$ that matches numerical results.

Utilizing the notation by Collins *et al.*, we begin with the spectral BTE in one dimension under the relaxation time approximation [23]:

$$\frac{\partial g_\omega}{\partial t} + \mu v_\omega \frac{\partial g_\omega}{\partial x} = \frac{g_0 - g_\omega}{\tau_\omega} \tag{1}$$

where $g_\omega$ is the phonon energy density per unit frequency interval per unit solid angle above the reference background energy, related to the distribution function and density of states as $g_\omega = \frac{\hbar \omega D(\omega)}{4\pi}(f_\omega - f_0(T_0))$. $\mu$ is the direction cosine, $v_\omega$ is the group velocity, $\tau_\omega$ is the relaxation time, and $g_0$ is the equilibrium energy density, given by $g_0 \approx \frac{1}{4\pi} C_\omega (T - T_0)$ in the linear response regime. In the TTG experiment, the temperature initially has a sinusoidal profile and in general obeys $T(x,t) = T_0 + h(t) T_m e^{iqx}$ in complex form where $T_m$ is the initial amplitude of the spatial variation, $q = 2\pi/\lambda$ is the grating wavevector, and $h(t)$ is the non-dimensional temperature that describes the decay of the initial temperature profile. Solving Eq. (1) and utilizing the equilibrium condition in the spectral case [32] to close the problem yields the integral equation for the non-dimensional temperature obtained previously [23,24]:

$$h(t) \int_0^{\omega_m} \frac{C_\omega}{\tau_\omega} d\omega = \int_0^{\omega_m} \frac{C_\omega}{\tau_\omega} b_\omega(t) d\omega + \int_0^t h(t') \int_0^{\omega_m} \frac{C_\omega}{\tau_\omega^2} b_\omega(t-t') d\omega \, dt' \tag{2}$$

where we have defined for simplicity $b_\omega(t) \equiv e^{-\frac{t}{\tau_\omega}} \text{sinc}(q v_\omega t)$. This integral equation is easily solved with a Laplace transform, and the temperature profile can be solved for with an inverse transform as obtained by Hua and Minnich [24]. Other methods to solving the BTE is to either obtain a numerical solution by solving the integral equation by finite differences [8,23], or by utilizing Monte Carlo techniques [31,33,34]. We depart from these established approaches by treating the unknown temperature distribution as a variational function and rewrite Eq. (2) by shifting all terms to one side of the equation to define:

$$H(t) = \int_0^{\omega_m} \frac{C_\omega}{\tau_\omega} b_\omega(t) d\omega - \bar{h}(t) \int_0^{\omega_m} \frac{C_\omega}{\tau_\omega} d\omega + \int_0^t \bar{h}(t') \int_0^{\omega_m} \frac{C_\omega}{\tau_\omega^2} b_\omega(t-t') d\omega dt' \quad (3)$$

If the function we guess for the temperature profile is the exact temperature profile that solves the BTE, then this function will be identically zero everywhere. The function $H$ has a physical interpretation as it has been defined from the integral equation which comes from the equilibrium condition of the BTE; it represents the error in energy conservation, and can be thought of as an artificial heat source/sink (up to constant factors such as $4\pi$ and $T_m$). In the exact case it should be zero everywhere, but since our trial function will not be the exact solution, we would like to optimize the function that makes $H(t)$ as close to zero as possible to minimize the error in our trial solution.

The optimization procedure can be done in several ways. One method is to mathematically define an error metric and minimize the error in order to calculate the variational parameter. Some common examples of error metrics are least squares $Er(\gamma) = \int_0^\infty H^2(t) dt$ and least absolute error $Er(\gamma) = \int_0^\infty |H(t)| dt$. Another approach is to require certain physical conditions to be met, and we can impose one physical condition for every variational parameter available in the trial solution. In this work, we use the following simple physical constraint to solve the variational parameter:

$$\int_0^\infty H(t) dt = 0 \quad (4)$$

Since *H* represents an artificial heat generation rate, the integral in Eq. (4) represents the total net heat generated over all time and solving it to be zero imposes the requirement that even if the energy conservation does not exactly hold at every moment in time as it does for the exact solution, it is satisfied over the entire decay time. The key point is to optimize the trial function, either by imposing physical conditions to be met or mathematical error functions to be minimized and there are various ways to do so.

Typically in the variational approach, one uses a trial function that is known from intuition about the system, and the trial function is optimized to minimize the chosen error function. In solving the BTE with this variational approach, the Fourier solution provides this trial function, especially since our goal is to extract the effective thermal conductivity (or in other cases properties such as interfacial thermal resistance, diffusivity, etc.) of the system.

For the one-dimensional TTG, the exact temperature solution of the diffusion equation is $T(x,t) = T_0 + T_m e^{iqx} e^{-\alpha q^2 t}$ where $\alpha$ is the thermal diffusivity. Therefore, we take for the trial function $\bar{h}(t) = e^{-\gamma t}$ with $\alpha_{eff} = \gamma / q^2$. The elegance of this approach is that it immediately utilizes the Fourier temperature field appropriately modified as an input, and optimizes to find the modified properties such as effective thermal conductivity that solves the BTE with minimized error, converting the difficult task of solving an integral equation for the temperature distribution into a simple task of performing integration. We note that in other heat transfer configurations, there can be multiple parameters in the trial solution, such as temperature slip that can occur due to the boundary resistance as well as effective thermal conductivity in the non-diffusive regime. Here, we demonstrate the technique on this simple case where only one variational parameter will be needed for simplicity.

Using the trial function, we can solve the condition of Eq. (4) to obtain for the thermal decay rate:

$$\gamma = \frac{\int_0^{\omega_m} d\omega \frac{C_\omega}{\tau_\omega}\left[1 - \frac{1}{\eta_\omega}\arctan(\eta_\omega)\right]}{\int_0^{\omega_m} d\omega C_\omega \frac{1}{\eta_\omega}\arctan(\eta_\omega)} \quad (5)$$

where we have defined the non-dimensional Knudsen number $\eta_\omega = q\Lambda_\omega = 2\pi\Lambda_\omega/\lambda$ where $\lambda$ is the grating period. As described, while different mathematical error metrics can be chosen or different physical conditions, we will show that the simple physical constraint of Eq. (4) does an excellent job in recovering the exact numerical results.

We utilize the normalized effective thermal conductivity for simplicity, defined as $\frac{k_{eff}}{k_{bulk}} = \frac{C\gamma}{q^2 k_{bulk}}$. The bulk thermal conductivity is given by $k_{bulk} = \frac{1}{3}\int_0^{\omega_m} C_\omega v_\omega \Lambda_\omega d\omega$. In Fig. 1, we compare results obtained for Si and PbSe, for which previous approaches for obtaining the effective thermal conductivity include assuming a constant MFP distribution [23], an exact numerical solution [23,24], and the two-fluid model [19]. The spectral numerical results are obtained by fitting the exact solution of Eq. (2), obtained by finite differences, to the Fourier exponential profile [23]. The spectral variational results are plotted from Eq. (5). The gray variational results are obtained by taking the gray limit of Eq. (5), extracting a gray suppression function, and inputting into the effective thermal conductivity, i.e. $k_{eff,gray} = \frac{1}{3}\int_0^{\omega_m} C_\omega v_\omega \Lambda_\omega S_{gray}(\eta_\omega) d\omega$. Note that this gray approximation is identical to the 'frequency integrated gray medium' approach performed numerically by Collins *et al.* [23]. Silicon is known to have a wide range of MFPs, and shows that the gray solution derived suppression function does a poor job in reproducing the exact numerical results. We also look at PbSe, which has a narrower range of MFPs. Note that the optimized solution agrees excellently with the exact numerical solution, which demonstrates the predictive power of the variational approach. Here the two-fluid model deviates from the exact solution at smaller grating periods. The gray suppression function performs better for this material due to its narrower range of MFP's as compared to silicon [23].

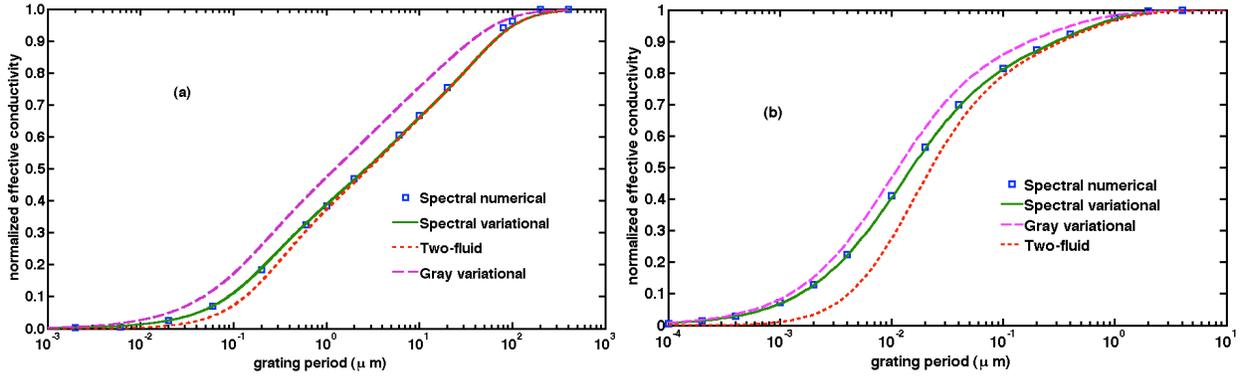

FIG. 1 Effective thermal conductivity of silicon (a) and PbSe (b). Here the effective thermal conductivity is plotted to compare the variational technique with the exact numerical technique and various approximations. The variational technique for the full spectral BTE demonstrates excellent agreement with the exact numerical solution.

From the definition of the effective thermal conductivity and the thermal decay rate of Eq. (5), we extract the suppression function:

$$S_\omega = \frac{\frac{3}{\eta_\omega^2}\left[1 - \frac{\arctan(\eta_\omega)}{\eta_\omega}\right]}{\int_0^{\omega_m} d\varpi \frac{C_\varpi}{C} \frac{\arctan(\eta_\varpi)}{\eta_\varpi}} \quad (6)$$

where $C$ is the heat capacity obtained by integrating the spectral heat capacity $C = \int_0^{\omega_m} C_\omega d\omega$. We note that although the numerator is dependent only on the ratio of MFP to the grating spacing and hence universal, the denominator depends in general on the material properties, This result is significant not only because it shows the suppression function is not universally dependent on a ratio of MFP to a length in the system, but also because we now have a way to properly address this problem analytically and more generally, numerically. The numerator is equal to the suppression function previously derived by Maznev *et al.* [19] and has been called the weakly quasiballistic suppression function [24]. Hua & Minnich have shown that there is in fact a correction to the suppression function in the full form, but their expression depends on the thermal decay time which intrinsically depends on the temperature solution to the BTE [24]. Our optimized solution provides the suppression function and illuminates its dependence on the grating period as well as its material property dependence. Furthermore, we can determine the validity domain of the two-fluid model by comparing the denominator of the optimized expression to unity. Thus, the following quantitative metric is obtained for the validity of the two-fluid model, $\left|\int_0^{\omega_m} d\omega \frac{C_\omega}{C} \frac{\arctan(\eta_\omega)}{\eta_\omega} - 1\right| \ll 1$. One could Taylor the suppression function of Eq. (6) expand for large values of the grating period relative to MFP to get an expression that is a first order correction to the two-fluid approximation for the thermal decay rate. In Fig. 2, we show the denominator of Eq. (6) for both Si and PbSe. We find that the two-fluid model can predict the effective thermal conductivity of Si with less than 5% error for grating spacings of 1 micron or higher. For PbSe, the two-fluid model has less than 5% error for grating spacings of 0.1 micron or higher. The cutoff grating spacing is larger for Si than for PbSe because Si has a MFP distribution that has a larger maximum MFP value than for PbSe, as shown in Fig. 1, hence demonstrates earlier deviation from the exact result.

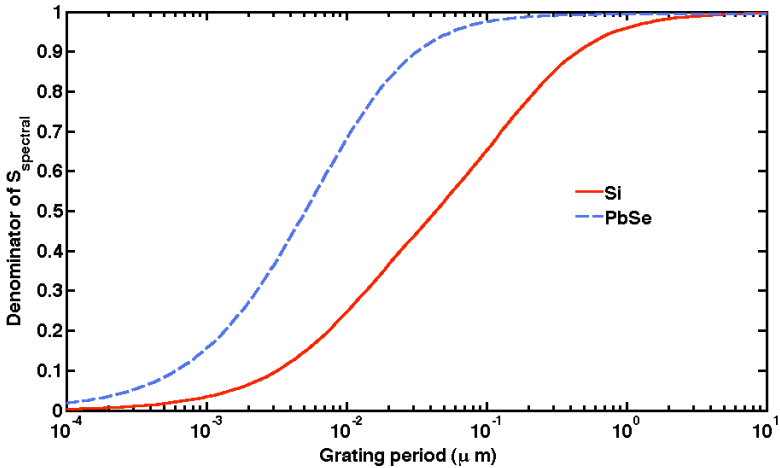

FIG. 2: Denominator of optimized spectral suppression function for Si (red) and PbSe (blue), yielding a metric for the validity of the two-fluid model.

The variational method can, of course, be applied to the gray case, for which we extract a suppression function that only depends on the Knudsen number. We take Eq. (6) and assume a constant MFP distribution to obtain:

$$S_{gray}(\eta_\omega) = \frac{3}{\eta_\omega^2}\left[1 - \frac{\arctan(\eta_\omega)}{\eta_\omega}\right]\left\{\frac{\eta_\omega}{\arctan(\eta_\omega)}\right\} \quad (7)$$

The term in the curly brackets is the additional factor we have obtained compared to the two-fluid model. The gray suppression function demonstrates a weaker suppression (higher effective thermal conductivity) than the two-fluid model due to this additional factor. The gray suppression function of Eq. (7) excellently reproduces the results of the normalized gray medium effective diffusivity obtained numerically previously by Collins *et al.* [23]. However, we have shown that indeed this approach of inputting the gray suppression function into the effective thermal conductivity expression is not universal, and performs rather poorly, especially for silicon as shown by the gray variational results from Fig.1.

In summary, we have developed a variational approach that yields a new way of extracting the effective thermal conductivity of the system by exploiting knowledge of the Fourier solution. In general, this approach to solving the temperature equation for the spectral BTE can directly yield the effective thermal conductivity from quasi-ballistic phonon transport experiments without brute force numerical solution of the BTE. We demonstrate the power of this approach by calculating the thermal decay rate as well as an analytical suppression function for one-dimensional transient grating experiments. Our spectral suppression function yields the exact suppression of thermal conductivity. We have shown that the suppression function is not universal, and utilizing the gray solution to the BTE does not perform well in reproducing the exact spectral data. Moreover, the variational approach developed here can be used as a universal technique for solving the BTE and obtaining both experimental observables, such as measured heat flux or thermal decay rate, as well as the effective thermal conductivity. This technique can be extended beyond the TTG problem, and can be used to calculate the effective thermal conductivities and suppression functions in different experimental geometries. This technique and our results here will allow for a better understanding of transport beyond the diffusive limit.

Acknowledgment: The authors would like to thank Dr. Kimberlee C. Collins for providing the material property data for Si and PbSe. This material is based upon work supported as part of the "Solid State Solar-Thermal Energy Conversion Center (S3TEC)", an Energy Frontier Research Center funded by the U.S. Department of Energy, Office of Science, Office of Basic Energy Sciences under Award Number: DE-SC0001299/DE-FG02-09ER46577.